\documentclass[letterpaper,twoside,journal,final,twosided]{IEEEtran}
\usepackage[T1]{fontenc}
\usepackage[latin9]{inputenc}
\usepackage{units}
\usepackage{amsmath}
\usepackage{amssymb}
\usepackage{graphicx}
\usepackage[unicode=true,pdfusetitle,
 bookmarks=true,bookmarksnumbered=false,bookmarksopen=false,
 breaklinks=false,pdfborder={0 0 1},backref=false,colorlinks=false]
 {hyperref}
\usepackage{breakurl}

\makeatletter

\newcommand{\lyxdot}{.}

\setkeys{Gin}{width=\linewidth}
\usepackage{cite}

\long\def\@makecaption#1#2{\ifx\@captype\@IEEEtablestring%
\footnotesize\begin{center}{\normalfont\footnotesize #1}\\
{\normalfont\footnotesize\scshape #2}\end{center}%
\@IEEEtablecaptionsepspace
\else
\@IEEEfigurecaptionsepspace
\setbox\@tempboxa\hbox{\normalfont\footnotesize {#1.}~~ #2}%
\ifdim \wd\@tempboxa >\hsize%
\setbox\@tempboxa\hbox{\normalfont\footnotesize {#1.}~~ }%
\parbox[t]{\hsize}{\normalfont\footnotesize \noindent\unhbox\@tempboxa#2}%
\else
\hbox to\hsize{\normalfont\footnotesize\hfil\box\@tempboxa\hfil}\fi\fi}
\ifCLASSOPTIONcompsoc
  \usepackage[caption=false,font=normalsize,labelfont=sf,textfont=sf]{subfig}
\else
  \usepackage[caption=false,font=footnotesize]{subfig}
\fi

\usepackage{dblfloatfix}

\@ifundefined{showcaptionsetup}{}{%
 \PassOptionsToPackage{caption=false}{subfig}}
\usepackage{subfig}
\makeatother

\begin{document}

\title{Innovative Telecommunications Training through~Flexible~Radio~Platforms}

\author{Ali Fatih Demir, Student Member, IEEE, Berker Peköz, Student Member,
IEEE, \\
Selçuk Köse, Member, IEEE, Hüseyin Arslan, Fellow, IEEE\vspace{-0mm}
\thanks{Manuscript received April 15, 2019; revised August 06, 2019;
accepted September 02, 2019.}\thanks{Ali Fatih Demir and Berker
Peköz are with the Department of Electrical Engineering, University
of South Florida, Tampa, FL 33620 USA (e-mail: afdemir@mail.usf.edu,
pekoz@mail.usf.edu).}\thanks{Selçuk Köse is with the Department
of Electrical and Computer Engineering, University of Rochester, Rochester,
NY 14627 USA (e-mail: skose@ece.rochester.edu).}\thanks{Hüseyin
Arslan is with the Department of Electrical Engineering, University
of South Florida, Tampa, FL 33620 USA, and also with the Department
of Electrical and Electronics Engineering, Istanbul Medipol University,
Istanbul 34810, Turkey (e-mail: arslan@usf.edu).}}
\maketitle
\begin{abstract}
The ever-changing telecommunication industry is in severe need of
a highly-skilled workforce to shape and deploy future generation communication
systems. This article presents an innovative telecommunication training
that is designed to satisfy this need. The training focuses on hardware
layers of the open systems interconnection model. It integrates theory,
numerical modeling, and hardware implementation to ensure a complete
and long-lasting understanding. The key telecommunication concepts
that are covered in the fundamental training phase are detailed along
with best teaching practices. In addition, the methods that enrich
the learning experience, such as gamified micro-tasks and interactive
use of daily telecommunication devices, are featured. The project
development case studies that cultivate creative thinking and scientific
interest are highlighted. Also, a well-established guideline to compose
the teaching environment that emphasizes hands-on experience is provided.
Therefore, the presented training can be exemplary to other institutions
that share the same mission to educate the distinguished engineers
of the future.
\end{abstract}


\markboth{IEEE Communications Magazine}{Demir \MakeLowercase{\textit{et al.}}: An Innovative Telecommunications Training Through Flexible Radio Platforms}

\section{Introduction}

\IEEEPARstart{T}{elecommunication} technologies have been evolving
from smoke signals to deep space high-definition video conferences
at an accelerating pace \cite{parkvall2017,consonni2010}. Such a
thriving technology requires a vast number of highly skilled engineers.
Therefore, a well-designed training is needed to satisfy the ever-changing
telecommunication industry\textquoteright s thirst for resilient professionals.
Distinguished telecommunication engineers must have a solid understanding
of the fundamental theory, an ability to translate this theoretical
knowledge to the numerical implementation, and capability of implementing
them in hardware. Also, they must be equipped with independent thinking
and creative problem-solving skills to pioneer future telecommunication
systems. Furthermore, these prodigious engineers have to express their
ideas clearly and function effectively on a team.

This article presents an innovative telecommunications training that
is designed to cultivate equipped telecommunication engineers by instilling
the following abilities:
\begin{itemize}
\item To identify, formulate, and solve telecommunication systems\textquoteright{}
problems by applying principles of mathematics, digital signal processing
(DSP), electromagnetic (EM) wave propagation, and communication theory.
\item To apply these fundamental concepts to design and test spectral/energy-efficient
advanced telecommunication systems considering a wide variety of system
scenarios, channel conditions, and hardware limitations.
\item To present designs and document outcomes taking a diverse audience
into account.
\end{itemize}
The training emphasizes hands-on experience and focuses on hardware
layers of the open systems interconnection (OSI) model \cite{OSI}.
At least a senior-level standing in electrical engineering is required
for the proposed training. Not only students but also industry professionals
who are looking to expand their understanding of telecommunications
are targeted. Although there exist several courses considering either
a particular technology \cite{padgett2006,cassara2006,linn2012} or
a teaching method \cite{aliakbarian2014,berber2018}, the proposed
training covers various telecommunication technologies and integrates
many modern teaching methods such as gamified micro-tasks and novel
doubly dispersive channel emulators. The foundations of the suggested
training were laid in \cite{guzelgoz-1-2010}, and the training is
rebuilt by including emerging concepts, contemporary teaching methods,
and futuristic projects over the last decade.

\section{Modern Teaching Methodologies}

\begin{figure}
\centering\includegraphics[width=1\columnwidth]{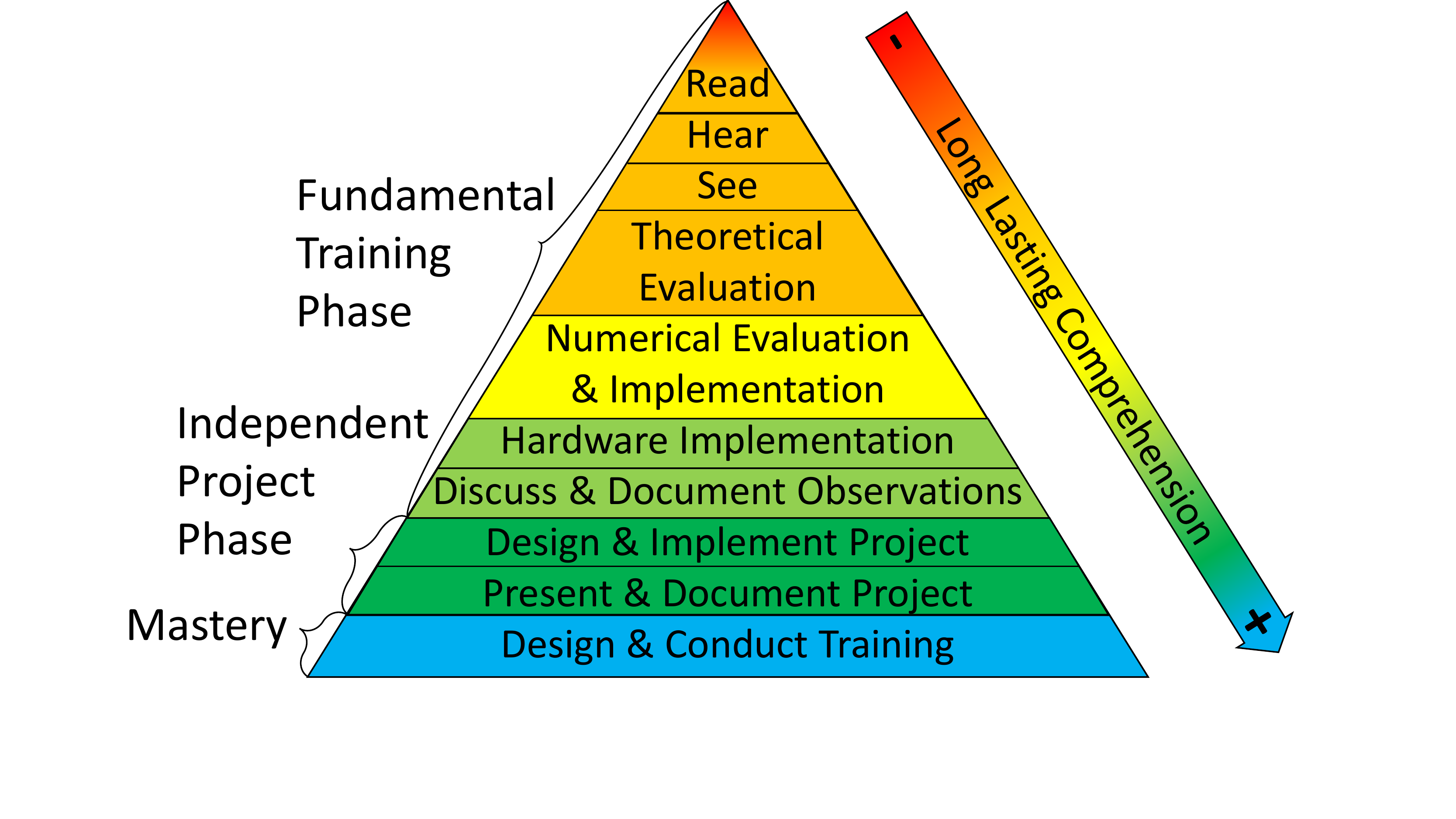}
\centering{}\caption{The cone of learning adapted from \cite{dale1969}. \label{fig:Experience-Cone}}
\end{figure}

Comprehending certain subjects such as telecommunications is very
challenging due to their highly abstract nature. The classical theoretical
training prevents observing the immediate relationship between cause
and effect. Also, theoretical knowledge without practice is destined
to perish. A complete understanding can be achieved by an interactive
learning experience \cite{dale1969,hoic-bozic2009} as shown in Fig.
\ref{fig:Experience-Cone}. The proposed telecommunication engineering
training is structured to modernize such a well-proven methodology.

A solid design builds on a good understanding of theoretical concepts.
However, telecommunication systems exhibit numerous non-linear behaviors
that make them difficult to model with closed-form expressions. Therefore,
numerical models are widely used to assist in the evaluation and visualization
of such complex systems. Trainees can manipulate the system and sub-system
parameters independently and grasp their individual effects using
various numerical tools. These skills are especially critical in their
professional career since troubleshooting problems with such tools
are easier than fixing them in the hardware prototyping stage. Nonetheless,
the validity of the aforementioned models is limited by the assumptions.
Therefore, theoretically and numerically verified designs must also
be implemented in hardware and tested under various channel conditions
and realistic scenarios. The hardware implementation can be achieved
practically using flexible radio platforms. Also, it should be pointed
out that the numerical and hardware implementations provide a convenient
way for trainees to design telecommunication systems by themselves
and assist reinforced learning.

The fundamental concepts of hardware layers are modularized and taught
separately to examine each sub-system rigorously. In light of the
discussions above, an ideal training should integrate theory, numerical
modeling, and hardware implementation to train future telecommunication
engineers well. Therefore, each training module starts with a theoretical
discussion, followed by numerical analysis, and is completed with
hardware implementation in the laboratory. The laboratory experiments
feature telecommunication devices that are used daily such as smartphones
and FM radio receivers to interest trainees. Furthermore, gamified
micro-tasks motivates them and establishes confidence in this challenging
field. The details of each module are given in the following section.
At the end of each module, trainees are expected to deliver a technical
report summarizing their key observations. These reports allow receiving
immediate feedback regarding trainees\textquoteright{} progress and
help them keep on track throughout the fundamental training phase.
Upon successful completion of this phase, trainees are examined, and
they proceed to the independent project development phase to demonstrate
their vast proficiency on the topic.

\section{Design of a Telecommunication Engineering Training}

\begin{figure*}
\centering\includegraphics[width=1\textwidth]{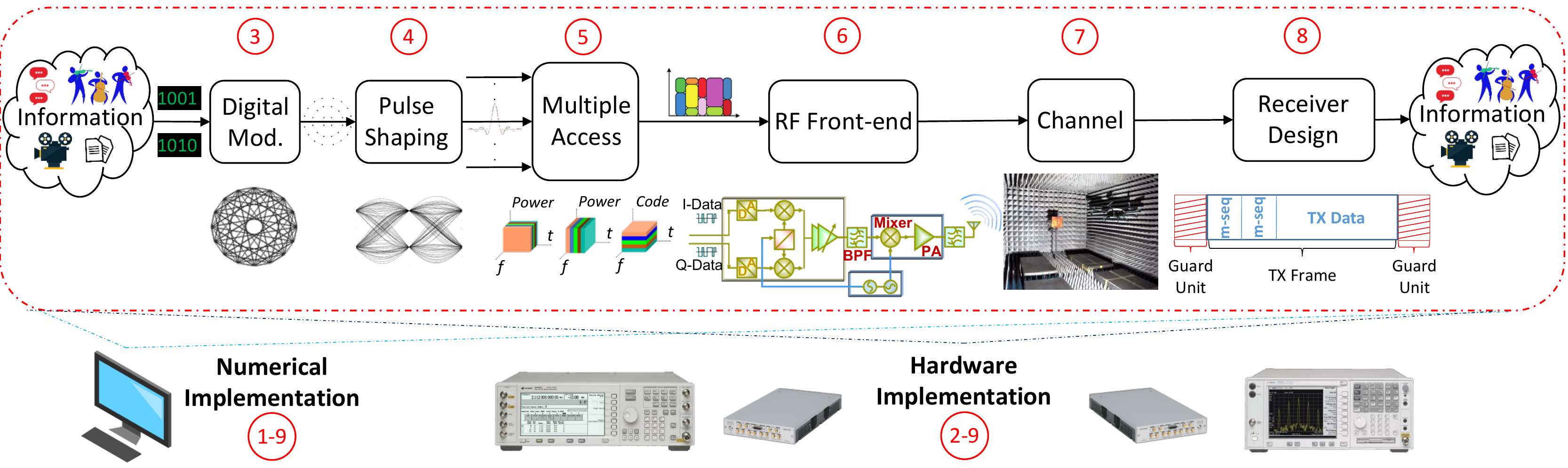}\caption{Functional model of the fundamental training. \label{fig:Functional-Model}}
\end{figure*}

The proposed training teaches telecommunications with an emphasis
on physical, data link, and network layers. The training content presents
the journey of bits through telecommunication systems as depicted
in Fig. \ref{fig:Functional-Model}. An instructor must compose the
teaching environment carefully to deliver this content effectively.
Once the teaching environment is established, the fundamental training
and independent project development phases can be conducted. The total
training duration depends on the audience. For example, the training
is offered to students in 15 weeks, whereas the training is delivered
to industry professionals on various extents. A good rule of thumb
would be to allocate $\nicefrac{2}{3}$ of the fundamental training
period for the independent project development phase. 

\subsection{The composition of a Telecommunication Teaching Environment}

\begin{figure}
\centering\includegraphics[width=1\columnwidth]{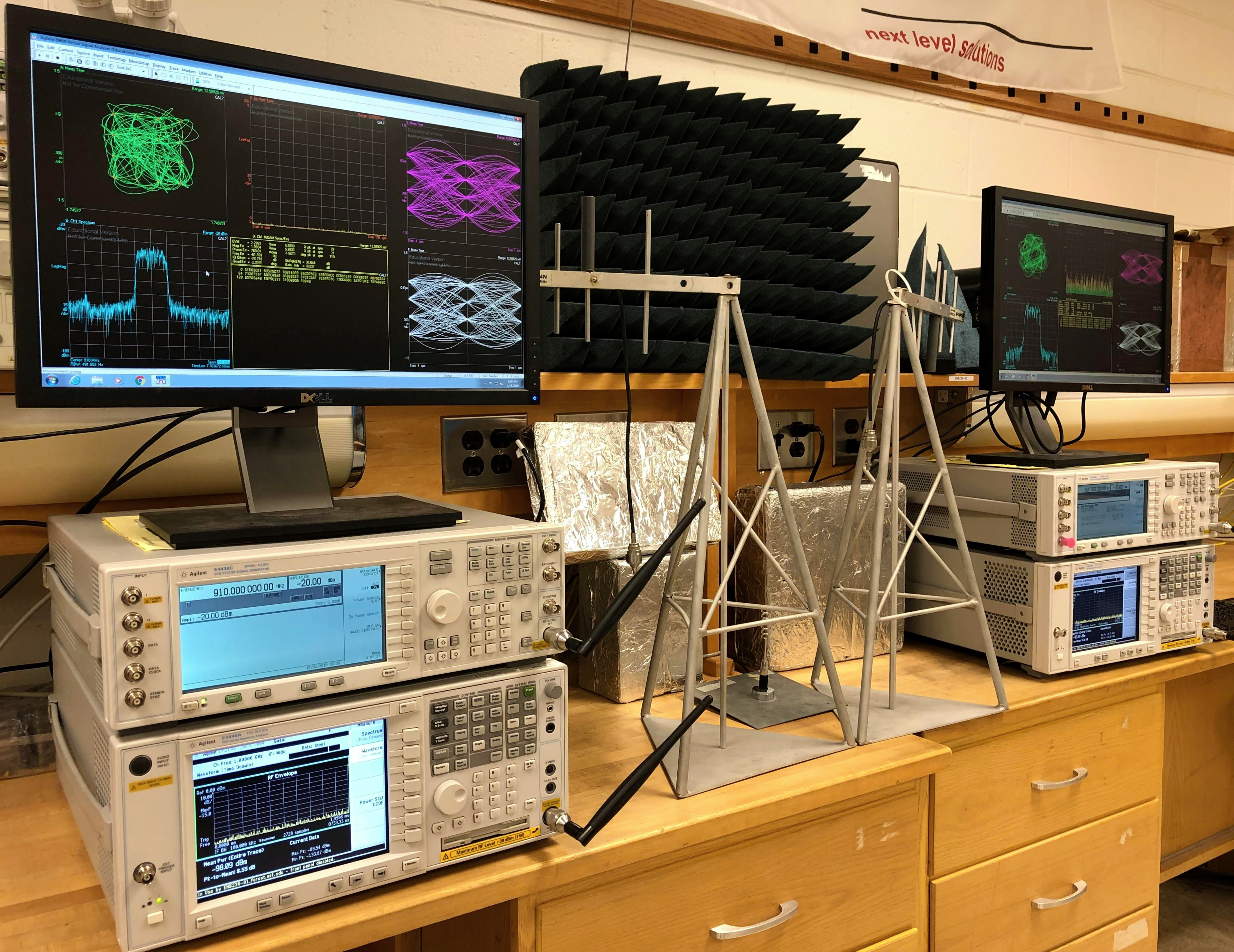}
\centering{}\caption{Basic SDR test-beds each consisting of a vector signal generator (VSG),
a vector signal analyzer (VSA), a computer, and cables/antennas. \label{fig:Test-bed}}
\end{figure}

The theoretical content and numerical modeling can be taught either
in a conventional computer laboratory or remotely. On the other hand,
the hardware implementation requires a flexible radio platform that
consists of software-defined radio (SDR) capable transmitter and receivers
along with the modular RF front-end and configurable channel emulators.
A basic SDR test-bed includes a vector signal generator (VSG), a vector
signal analyzer (VSA), and a computer that runs telecommunication
system design and analysis tools as well as cables/antennas as shown
in Fig. \ref{fig:Test-bed}. A VSG can generate signals using various
digital modulation techniques and baseband pulse shapes. Also, it
can multiplex them in various domains such as time, frequency, and
code to obtain standard and custom waveforms. These waveforms can
either be generated internally using the VSG in standalone operation
or externally using the computer, and are conveyed to VSAs. A VSA
has the ability to demodulate the standard and custom signals in a
standalone mode similar to a VSG. Also, it can convey the in-phase
and quadrature (IQ) samples of the received signal to a computer for
processing. The interaction between a computer and VSGs/VSAs is an
excellent mechanism for teaching, studying, and analyzing the current
and upcoming telecommunication systems.

\begin{figure}
\centering\includegraphics[width=1\columnwidth]{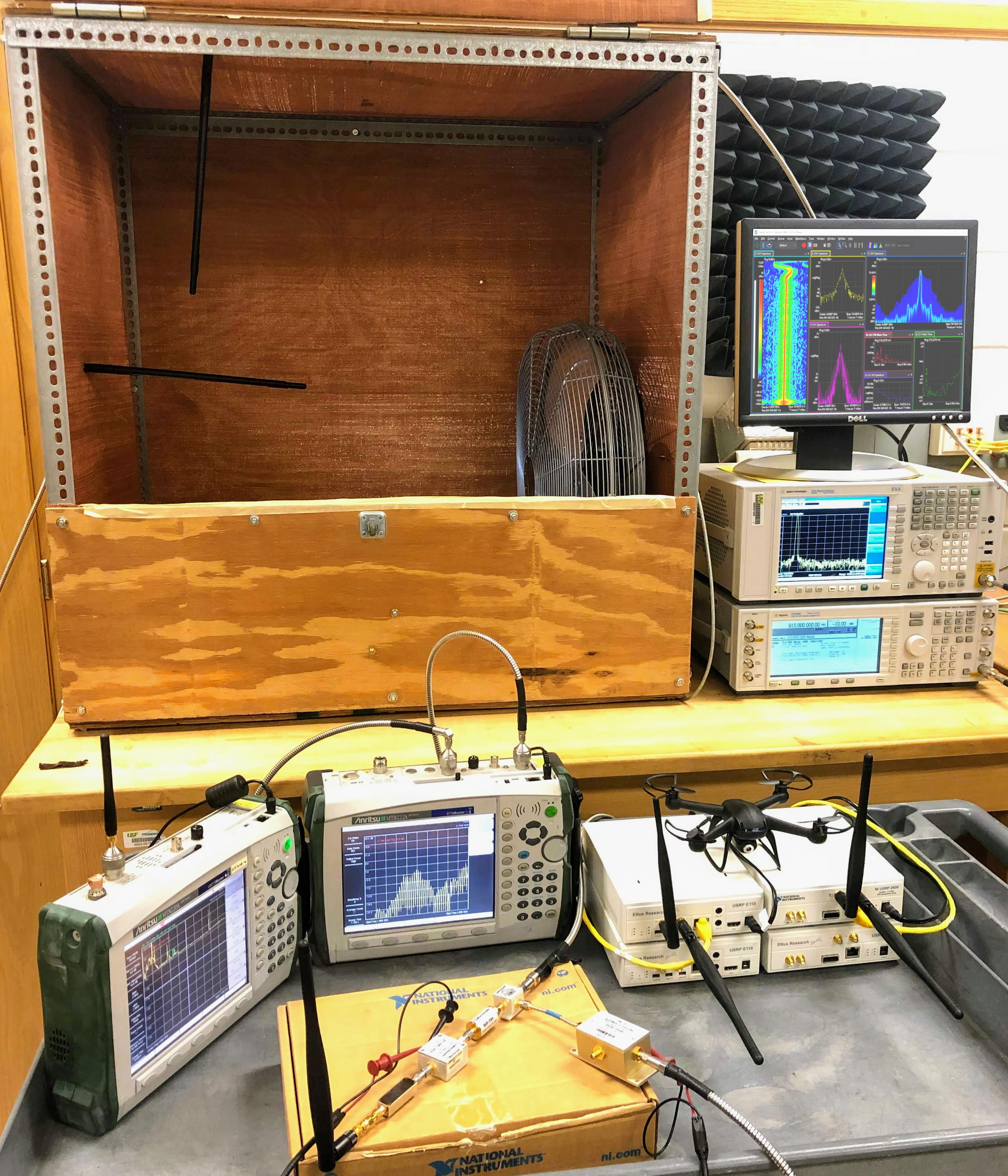}
\centering{}\caption{A flexible radio platform that features mobile SDR equipments, modular
RF front-end, and configurable channel emulators. \label{fig:Equipments}}
\end{figure}

The system design, scenarios, and channel conditions can be enriched
further by extending the test-bed with mobile transceivers, modular
RF front-end, and channel emulation tools as presented in Fig. \ref{fig:Equipments}.
IQ modems, digital to analog converters (DACs), analog to digital
converters (ADCs), mixers, voltage-controlled oscillators (VCOs),
power amplifiers (PAs), band-pass filters (BPFs), cables, and antennas
are some of the critical RF front-end components of telecommunication
systems. Testing and measuring a telecommunication system using a
modular RF front-end provides the ability to analyze the function
and effect of each component separately. In addition, mobile transceivers
allow emulating various scenarios such as cellular hand-off and positioning.
For high accuracy requirements, handheld VSGs and VSAs are preferable,
whereas an abundance of low-cost SDR hardware is more convenient for
experienced trainees to emulate applications involving networking
and interference management. Furthermore, diverse channel conditions
can be mimicked via reverberation and anechoic chambers, unmanned
aerial vehicles (UAVs), and metal fans \cite{guzelgoz-2-2010}. The
frequency selectivity of the channel is controlled through the use
of chambers, whereas time selectivity is managed using multi-speed
fans.

\subsection{Fundamental Training Phase}

Basic telecommunication concepts can be taught in eight modules as
numbered in Fig. \ref{fig:Functional-Model}. Also, the final module
adapts trainees to contemporary telecommunication systems. A detailed
description is provided in the training website\footnote{http://wcsp.eng.usf.edu/courses/wcsl.html},
and the key components of these modules are summarized as follows:

\subsubsection{Introduction to the Basic Telecommunication Concepts}

In this module, the theoretical aspects of the concepts that will
be covered throughout the training are introduced. Also, a transceiver
is modeled numerically to present the complete picture of a telecommunication
system. First, text messages for each test-bed are encoded to bits,
modulated, pulse-shaped, and multiplexed into different bands in the
presence of additive white Gaussian noise by the instructor. The waveforms
are supplied to trainees with combinations of various RF impairment
(i.e., frequency offset, phase noise, and PA nonlinearities) levels
and signal to noise ratios (SNRs). Trainees are required to down-convert
the signals and develop baseband receiver algorithms to complete the
gamified microtask of this module, which is detecting individualized
transmitted messages. Also, trainees assess the performance through
the observation of bit error rate (BER) and several diagrams such
as constellation, IQ polar, and eye. The benefit of this module is
twofold. First, trainees get familiar with the identification of telecommunication
systems' problems. Second, trainees become acquainted with utilizing
numerical tools to design and analyze telecommunication systems.

\subsubsection{Introduction to the Basic SDR Test-bed}

The primary objective of this module is to familiarize trainees with
a basic SDR test-bed. Trainees generate standard single-carrier and
multi-carrier waveforms at the VSGs and assess their performances
using the built-in functions of the VSAs along with the corresponding
computer software. The concepts that are theoretically and numerically
covered in the first module are implemented in the hardware. The instructor
allocates a different portion of the ISM band to each test-bed, and
trainees practice transceiving custom telecommunication signals. Not
only the signals that are generated by VSGs but also over the air
signals such as FM, Wi-Fi, and PCS are analyzed in this module. Observing
the FM radio spectrum interests trainees and relates the taught concepts
to daily life. Also, the burst transmission structure in ISM and PCS
bands upon Wi-Fi and cellular calls is demonstrated. In the last part
of the module, the received signal at VSA is downloaded to a computer,
and spectro-temporal characteristics are analyzed using numerical
tools. This part is especially important to illustrate the interconnection
between the components of a basic SDR test-bed.

\subsubsection{Digital Modulation}

Trainees commence the telecommunication system design by mapping bits
to symbols after the overview in the first two modules. The theory
part of the module teaches the purpose of digital modulation and its
limiting factors such as SNR and peak-to-average power ratio (PAPR).
In the numerical and hardware implementation parts, various modulation
schemes are demonstrated using built-in functions and characterized
in terms of power efficiency, spectral efficiency, and ease of implementation.
The analyses start with the throughput comparison of BPSK and higher-order
QAM and PSK modulation schemes as a function of SNR to demonstrate
the trade-off between spectral and energy efficiency. This trade-off
is made clear through the observation of error vector magnitude (EVM),
IQ polar diagram, eye diagram, and power CCDF curve. After initial
characterization, alternative modulation schemes such as $\pi$/2-BPSK,
$\pi$/4-DQPSK, OQPSK, and MSK that enhance PAPR to overcome hardware
limitations at the expense of implementation complexity are demonstrated
and examined. After trainees comprehend the distinguishing characteristics
of various digital modulation schemes, the instructor transmits a
signal, and trainees attempt to figure out the digital modulation
scheme blindly. Trainees understand the motivation behind the existence
of various digital modulation schemes and link adaptation to channel
conditions in this module.

\subsubsection{Baseband Pulse Shaping}

The objective of this module is to study the impact of baseband pulse
shaping filters, which map digitally modulated symbols into waveforms,
on telecommunication system performance. The discussions take off
with orthogonal raised-cosine (RC) filters. Trainees alter the roll-off
parameter that controls the spectro-temporal characteristics, and
they observe changes in the spectrum, time envelope, power CCDF curve,
IQ polar diagram, and eye diagram. Furthermore, the performance of
the RC pulse shaping filter at the transmitter side is compared with
using root-raised-cosine (RRC) pulse shaping filter at both transmitter
and receiver, and superiority of matched filtering is demonstrated.
Also, an RRC pulse-shaped signal is transmitted without matched filtering
at the receiver, and the presence of inter-symbol interference is
pointed out. Once trainees are accustomed to the non-orthogonality
in the filter design, the discussion continues with Gaussian pulse
shaping filters that are preferred due to their spectral confinement.
The similarity between the bandwidth-time (BT) product of Gaussian
pulse shaping filters and the roll-off factor of RC pulse shaping
filters are pointed out. GSM signals with different BT products are
generated, and their performances are evaluated as similar to RC pulse-shaped
signals. After understanding the spectro-temporal characteristics
of baseband pulse shaping filters, the instructor transmits signals,
and trainees predict the filter parameters. The fundamentals for spectral-
and energy-efficient design are taught in this module, and robust
filter design against RF front-end impairments and multiple access
wireless channel are further elaborated in the following modules.

\subsubsection{Multiple Accessing}

The purpose of this module is to acquaint trainees with resource allocation
among multiple users and to manage resulting interference. Previously,
the test-beds share the channel in an FDMA manner. In this module,
they get familiarized with other multiple accessing schemes such as
TDMA and CDMA. Initially, GSM signals with varying relative burst
powers for each slot are broadcasted by the instructor. Each slot
is assigned to a certain test-bed along with associated relative burst
power. Trainees observe the TDMA frame structure and locate their
slot. In the following stage, the instructor transmits various DSSS-CDMA
signals by assigning different codes to each test-bed. As the number
of active codes increases, trainees observe the power CCDF curve and
conclude that an increased number of random variables degrades PAPR
characteristics. Also, a joint time-frequency analysis is performed
on Bluetooth signals from two smartphones using the spectrogram. The
smartphones\textquoteright{} distances to a VSA are utilized to identify
the hopping structure of each frequency hopping spread spectrum code.

Upon introducing conventional resource allocation techniques, multiple
access interference is explained. Adjacent channel interference (ACI)
is demonstrated by transmitting equipower RRC pulse-shaped signals
in adjacent bands simultaneously without a guard band in between.
Although the test-beds can demodulate their signals with slight performance
degradation, they cannot demodulate each others\textquoteright{} signals
properly due to the near/far effect. In this step, trainees utilize
their knowledge from the previous module and adjust the RRC roll-off
factor to mitigate ACI. In the next stage, test-beds generate single-carrier
signals at the same frequency and emulate a co-channel interference
scenario. They observed that their communication performance significantly
degrades since there is an insufficient spatial separation between
the test-beds. Afterward, one of the test-beds switches to the DSSS
scheme, and it is demonstrated that spread spectrum systems are more
robust against narrowband interference. This module integrates previously
learned concepts and improves multi-dimensional signal analysis skills. 

\subsubsection{RF Front-end}

Teaching the functionalities and characteristics of critical RF front-end
components is the core objective of this experiment. In the first
stage of this module, the instructor informs trainees on IQ modems
and familiarize them with various impairments using the internal IQ
modulator on VSGs. IQ gain imbalance, quadrature offset, and DC offset
are intentionally introduced, and their effect on IQ polar diagram,
eye diagram, and EVM are exhibited. Following this stage, quantization
effects are demonstrated by connecting an external DAC with various
resolutions to VSGs and altering the dynamic range of VSAs. Trainees
observe the results of quantization errors such as spectral regrowth,
increase in EVM, interpolation issues in eye and IQ polar diagrams.
Afterward, the analog signal is conveyed to an upconverter unit that
consists of a mixer and a VCO to shift the baseband signal to IF or
RF. Trainees compare the stability of low-end standalone VCOs\textquoteright{}
output with that of VSGs. The reference signals generated by both
VCOs are passed through mixers for upconversion. It is demonstrated
that low-end standalone VCOs cannot output stable tones and produce
phase noise. 

Trainees analyze the effect of nonlinear behavior of mixers and PAs.
First, they observe the spectrum while operating the PA in the linear
region. Hence, the harmonics generated only by the mixer are presented.
Afterward, the transmitted power is gradually increased, and power
CCDF curve, EVM, IQ polar diagram, and spectrum are observed. It is
demonstrated that increasing the transmit power beyond the hardware
limitations does not increase SNR. BPFs are provided to suppress harmonics
caused by both components. Furthermore, PAPR is reduced by using digital
modulation techniques that omit zero-crossings and utilizing RRC pulse
shapes with a higher roll-off factor. Thus, trainees understand both
the inconveniences of these devices and techniques to cope with their
disadvantages. Also, this module provides them an opportunity for
building a complete RF front-end by connecting all the aforementioned
components.

\subsubsection{Propagation Channel}

The purpose of this module is to present fundamental EM wave propagation
concepts. The channel imposes various effects on EM waves such as
distance and frequency-dependent path loss, large scale fading (i.e.,
shadowing), and small scale fading (i.e., delay spread and Doppler
spread). In the first stage, the distance- and frequency-dependent
characteristics of the path loss are depicted by measuring the signal
with different antenna separations at different frequencies. Trainees
sweep the FM band and record the received power levels of various
stations. They obtain the transmit power, antenna height, and distance
information online. Afterward, a simple statistical path loss model
is derived considering the measurements from the stations with the
same parameters. Also, shadowing is demonstrated by pointing out the
variation around the path loss.

Small scale fading can originate from two phenomena; namely, delay
spread and Doppler spread. In the second stage, a reverberation chamber
is used to control the delay spread amount in an indoor environment.
Trainees alter the transmission bandwidth, and they observe a frequency
selective channel when the signal bandwidth exceeds the coherence
bandwidth of the channel. It is challenging to demonstrate mobility
in an indoor environment. Either the transceiver or the environment
must be mobilized. A multi-speed metal fan is utilized to create the
variation in time. Trainees place a metal fan in between two antennas
and operate it at different speeds while monitoring the spectrogram.
Furthermore, a doubly-dispersive channel is emulated by moving the
fan inside the reverberation chamber \cite{guzelgoz-2-2010,kihero2019}.
Finally, trainees apply the concepts that are discussed in the previous
modules to design telecommunication systems robust against channel
effects. For example, the symbol rate is decreased to elevate immunity
to delay spread, whereas it is increased to boost resistance to Doppler
spread. Furthermore, the RRC roll-off factor is raised to improve
the performance in both spreads with a penalty of degraded spectral
localization.

\subsubsection{Receiver Design}

The objective of this module is designing a complete digital baseband
receiver. The instructor transmits a BPSK modulated TDMA frame. The
frame consists of individualized messages following different repeated
m-sequences for each test-bed. Trainees capture the signal and record
it for offline processing. They start with the coarse time synchronization
step to pinpoint the position of their desired message. Since two
m-sequences are appended back to back, their autocorrelation gives
a rough estimate of the allocated slot beginning. Upon the coarse
time synchronization step, the frequency offset amount is estimated
using the constant delay between the consecutive m-sequences. The
estimated frequency offset is compensated for the frequency synchronization.
Then, fine time synchronization is performed by cross-correlating
the local m-sequence copy with the frequency offset compensated signal.
Trainees downsample the signal at the best sampling locations using
the fine time synchronization information. Afterward, channel estimation
must be carried out to compensate for the channel effect. The m-sequence
of the received signal is compared with the local m-sequence copy
in order to estimate the wireless channel. The phase and amplitude
of the received signal are corrected upon estimation. In the final
stage, digital demodulation is performed, and estimated bits are mapped
to characters to reveal the transmitted information. Trainees are
thrilled to be able to demodulate a physical signal similar to their
daily devices. This module finalizes the modules that are pointed
out in the functional diagram on Fig. \ref{fig:Functional-Model}
and demonstrates a complete picture of a basic single-carrier communication
system.

\subsubsection{OFDM}

The main features of OFDM modulation are studied in the last training
module. The instructor supplies trainees with a cyclic prefix (CP)-OFDM
transmission script along with the VSG and VSA connection libraries.
In the first stage, trainees are expected to develop their OFDM baseband
receiver algorithms similar to the previous module. Since there is
no m-sequence in the frame, CP is used for time and frequency synchronization.
Trainees repeat the hardware implementation steps of RF front-end
and propagation channel modules for multi-carrier communication after
the receiver design. FFT size, number of active subcarriers, CP length,
and modulation order are altered in each step, and their effect on
the telecommunication system performance is analyzed through power
CCDF curve, constellation, and spectrum. Also, they compare the impairments
for single-carrier and multi-carrier communication schemes and point
out the main differences. This module combines all the concepts that
are covered in this training within the perspective of an advanced
modulation scheme that is used in current cellular and Wi-Fi systems.

\begin{figure*}
\centering\subfloat[\label{fig:PIC-FSK}]{\includegraphics[width=0.33\textwidth]{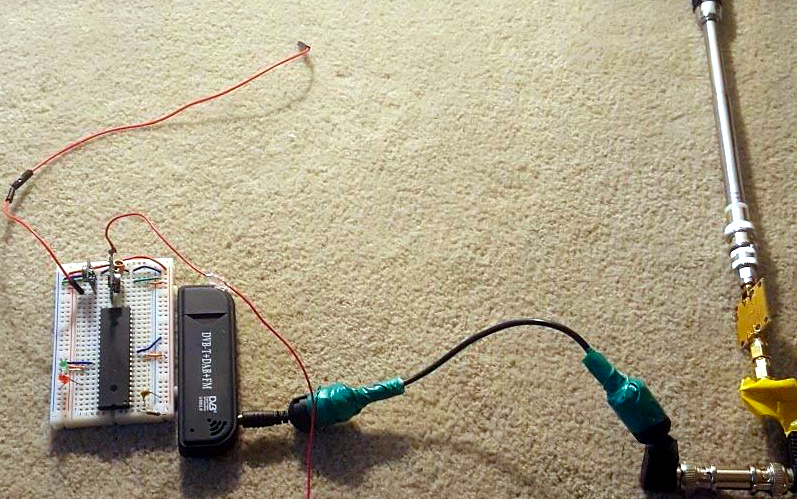}

}\subfloat[\label{fig:InVivo}]{\includegraphics[width=0.33\textwidth]{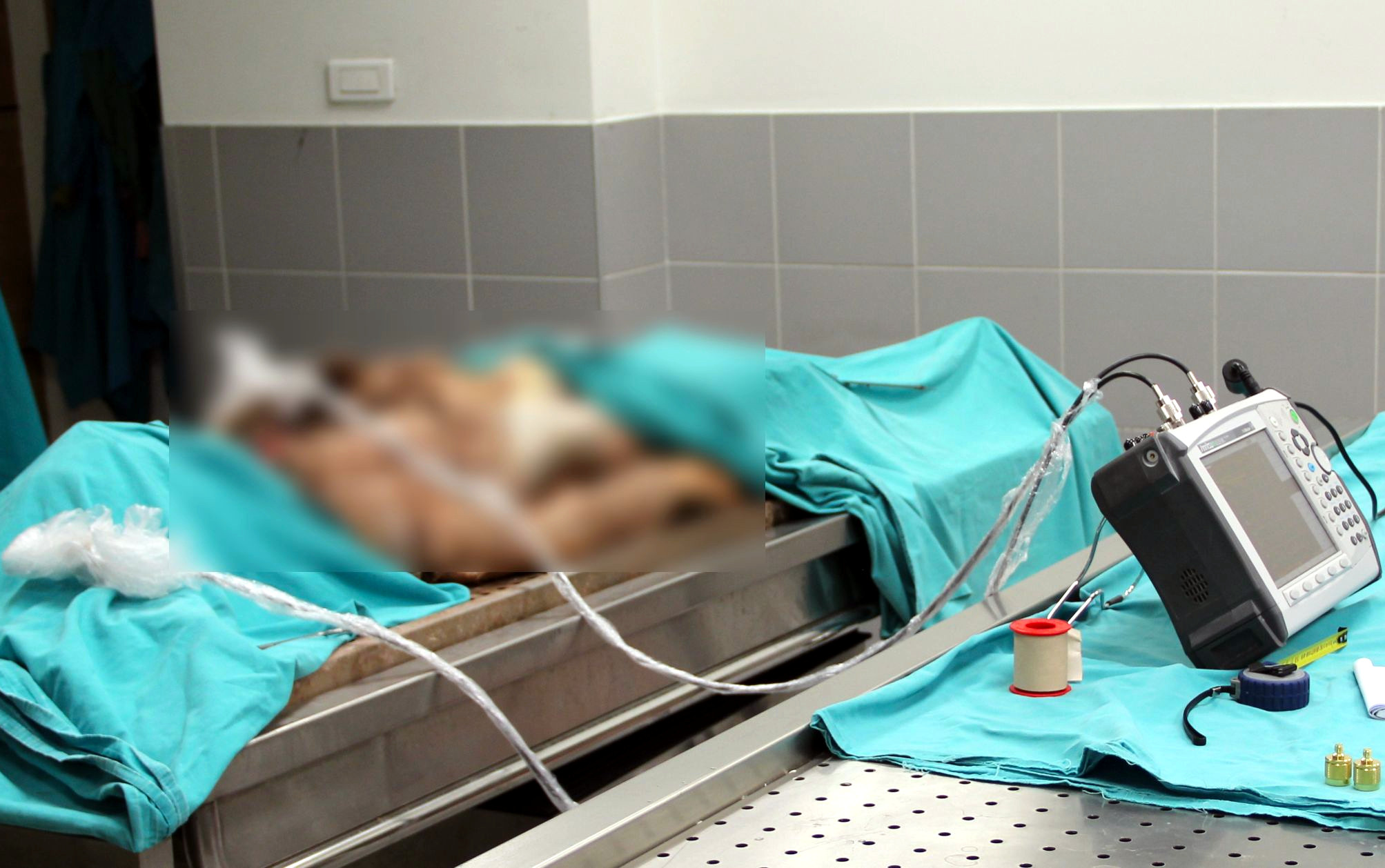}

}\subfloat[\label{fig:Object-ID}]{\includegraphics[width=0.33\textwidth]{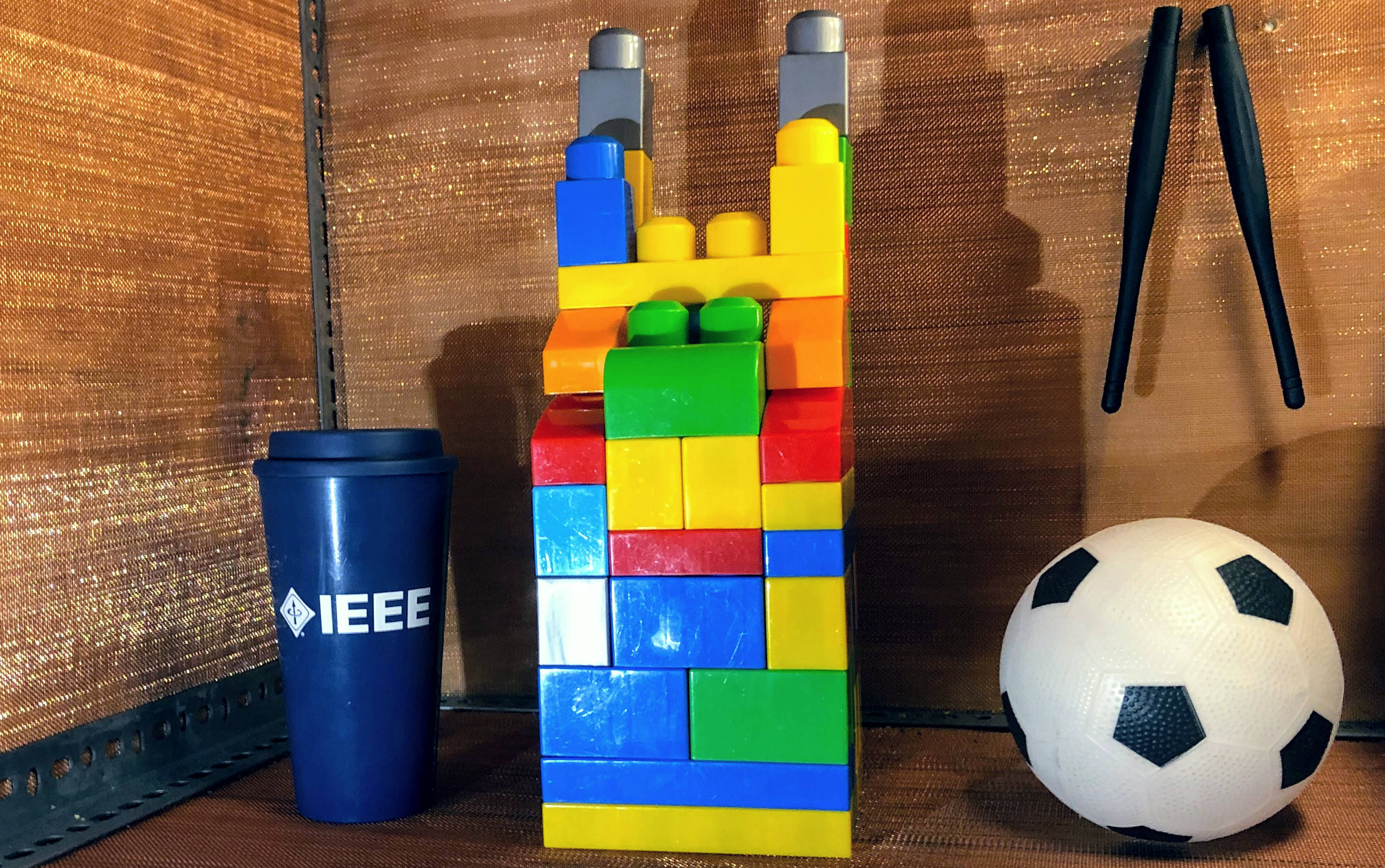}

}

\centering\subfloat[\label{fig:Ang-Est}]{\includegraphics[width=0.33\textwidth]{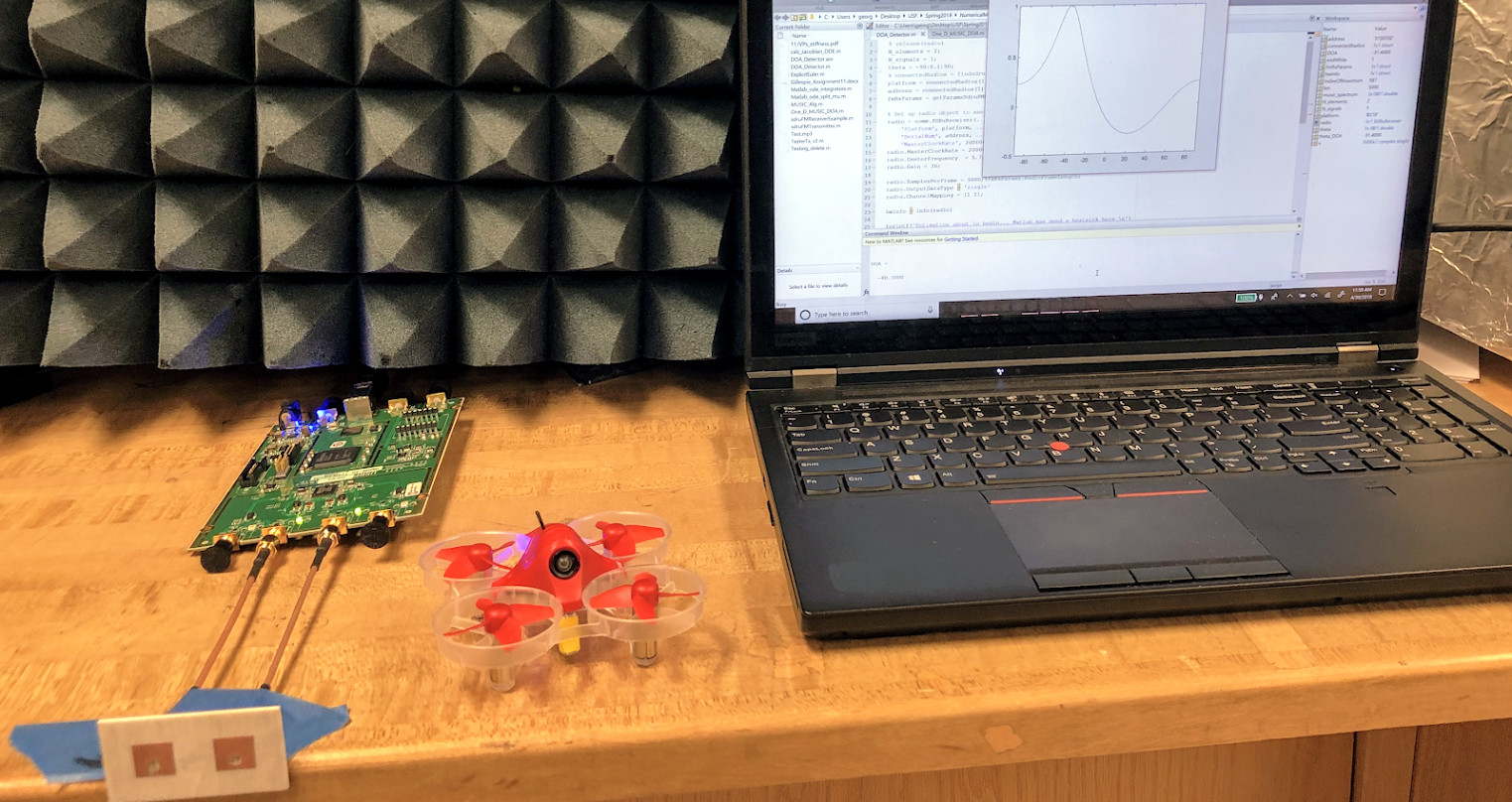}

}\subfloat[\label{fig:Beamformer}]{\includegraphics[width=0.33\textwidth]{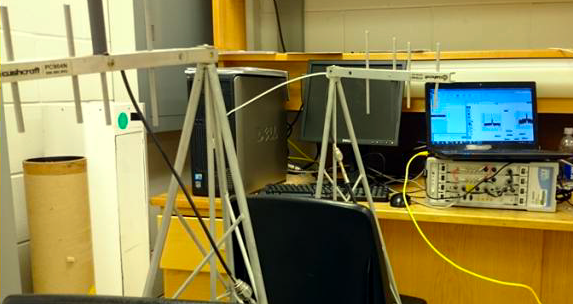}

}\subfloat[\label{fig:mmW}]{\includegraphics[width=0.33\textwidth]{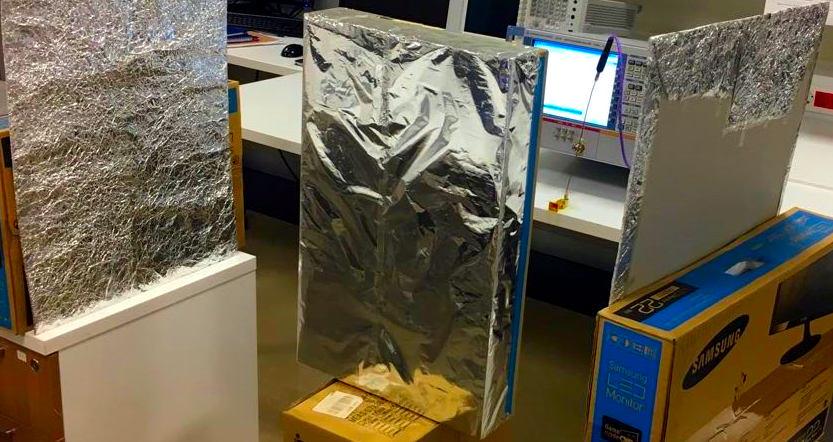}

}

\caption{Sample projects: (a) PIC driven SDR; (b) \emph{in vivo} channel characterization
\cite{demir2017}; (c) object identification through channel features;
(d) angle-of-arrival estimator for UAVs using a MIMO capable low-cost
SDR; (e) MIMO front-end design; (f) adaptive beamformer for mmW \cite{dogan2018}.\label{fig:Sample-Projects}}
\end{figure*}

\subsection{Independent Project Development Phase}

Trainees become ready to pursue independent projects upon the completion
of the fundamental training phase. Although novelty is not enforced,
it is highly encouraged. Considering the last 50+ projects that are
completed under the authors\textquoteright{} supervision, the feasible
and educational projects are categorically summarized to inspire prospective
trainees as follows:

\subsubsection{Software-defined radio (SDR)}

The expertise acquired with the higher-end SDR platforms in the fundamental
training phase equips trainees with the ability to build their low-cost
SDRs. These projects are especially suitable for embedded system developers
seeking to implement the Internet of Things transceivers. Fig. \ref{fig:PIC-FSK}
depicts an exemplary PIC driven FSK transmitter with its RF front-end.
The budget-limited SDR hardware comes with the price of excessive
RF front-end impairments. Trainees must also develop advanced baseband
algorithms to mitigate IQ impairments, PA nonlinearities, and low-resolution
DAC/ADC quantization issues.

\subsubsection{Wireless channel}

The fundamental training phase provides the ability to comprehend
basic features of conventional cellular and WiFi channels. Trainees
with strong propagation engineering backgrounds may characterize and
model different environments such as the promising in body channel
(Fig. \ref{fig:InVivo} \cite{demir2017}). Furthermore, advanced
channel features can be extracted to determine line-of-sight/non-line-of-sight
conditions or to design an object identifier via machine learning
(Fig. \ref{fig:Object-ID}). Location services driven by channel properties
are also compelling projects. Triangular positioning, channel-based
authentication, distance and angle-of-arrival estimation (Fig. \ref{fig:Ang-Est})
techniques are exemplary candidates. Moreover, the novel doubly-dispersive
channel emulators \cite{guzelgoz-2-2010,kihero2019} used in the fundamental
training phase were originally designed and implemented as trainee
projects.

\subsubsection{Wireless channel counteractions}

After obtaining a well-understanding in channel characteristics, trainees
can exploit them to improve telecommunication systems further. For
example, MIMO systems are designed to take advantage of spatial diversity.
Trainees with antenna and RF circuit design experience prefer MIMO
front-end design and manufacturing (Fig. \ref{fig:Beamformer}), whereas
trainees with solid communication theory background favor MIMO power
allocation, channel estimation, and compensation projects. Furthermore,
the mmW channel, which will be deployed in 5G, can be studied as well.
For instance, the mmW blockage issue is dealt with an adaptive antenna
beamwidth implementation as shown in Fig. \ref{fig:mmW} \cite{dogan2018}.

\subsubsection{Signal intelligence}

Trainees with DSP and communication theory backgrounds may utilize
their extensive knowledge on multi-dimensional signal analysis, and
communication channel to pursue signal intelligence applications.
These applications include non-data aided blind receivers which estimate
various parameters such as symbol rate, modulation type, and pulse
shaping filter parameters in the presence of RF front-end and channel
impairments. Furthermore, trainees can conceal the transmitted signal
to prevent eavesdropping using various single- and multi-antenna physical
layer security techniques such as artificial noise transmission. Also,
trainees may utilize channel-based authentication methods to identify
the legitimate transmitter as well.

\subsubsection{Multiple access and interference management}

The fundamental training phase mostly covers the centralized and orthogonal
multiple accessing schemes. Alternative schemes such as cognitive
radio (CR) and non-orthogonal multiple access (NOMA) attract trainees
with data link layer interest and cultivate numerous projects. For
example, the spectrum is utilized opportunistically in CR, and trainees
avoid interfering the primary user. Furthermore, non-contiguous frequency
resources can be aggregated in these projects. NOMA is another spectral-efficient
resource utilization scheme and can be realized with multi-user detection,
blind source separation, and interference cancellation algorithms.
Moreover, trainees may manage self-interference and develop full-duplex
communication schemes to improve the capacity of telecommunication
systems.

\subsubsection{Standards}

Trainees can improve their expertise in telecommunication systems
by partially implementing hardware layers of various standards such
as 3G/4G/5G cellular, IEEE 802.11, Bluetooth, and stereo FM. In addition,
those interested in network layer may realize vertical handoff mechanisms
to switch between different standards in a heterogeneous network scenario.
Mobile SDR equipment is essential in these projects. Moreover, trainees
can asses new technologies for future standards. For example, various
waveforms might be implemented to evaluate their performances in realistic
scenarios considering diverse channel conditions, multiple access
interference, and RF front-end impairments.

\section{Trainee Improvement}

The fundamental training phase, which integrates theory, numerical
modeling, and hardware implementation, leads to competent telecommunication
engineers with a solid education. Trainees who completed the training
are distinguished in their further telecommunication education and
career. Their academic success was tracked, and their superiority
to those who had not completed this training is revealed in both \cite[Table V]{guzelgoz-1-2010}
and Table \ref{tab:Academic-Success-Comparison}. Student final scores
in telecommunications-related courses from 2016 to 2019 are analyzed.
Seventy students who have taken this training succeed better in the
listed courses than 330 students who have not as presented in Table
\ref{tab:Academic-Success-Comparison}. Also, \cite[Table IV]{guzelgoz-1-2010}
points out that they feel more confident with telecommunication system
design and analysis after the training.

The independent project development phase requires trainees to acquire
and apply new knowledge as needed without well-defined instructions
and sharpens their analysis and synthesis skills. Also, it promotes
teamwork and generates synergistic engineers. Trainees improve their
presentation skills by demonstrating their projects effectively to
a wide range of audience that includes undergraduate students, graduate
students, faculty, and industry professionals. Discussions that take
place during presentations and feedback received afterward are reportedly
beneficial for professional interviews. The documentation of projects
benefits technical writing skills. Furthermore, notable projects are
encouraged for publication, and scientific interest among trainees
is cultivated.

Finally, the instructors who are assisting trainees throughout their
telecommunication education journey also benefit from tutoring, which
completes their mastery as pointed out in Fig. \ref{fig:Experience-Cone}.

\begin{table}[t]
\caption{Academic Success Comparison \label{tab:Academic-Success-Comparison}}

\centering\includegraphics[width=1\columnwidth]{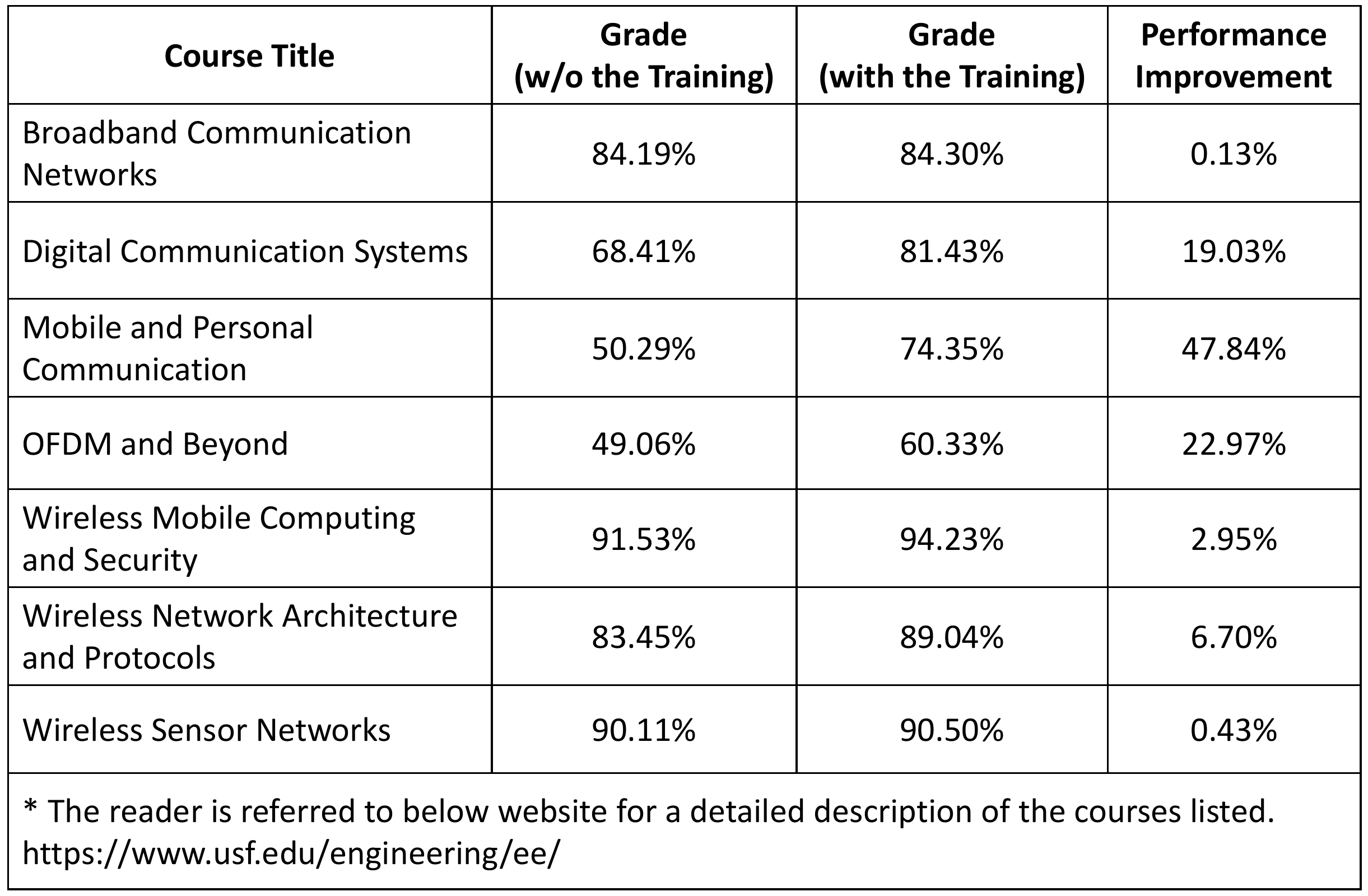}
\end{table}

\section{Conclusions}

The industry and academia require well-trained telecommunication engineers.
The proposed training is a well-established guideline to educate outstanding
engineers and to build a bridge between these two institutions by
integrating theoretical proficiency, numerical modeling skills, and
hands-on experience. The provided array of skills as well as professional
qualities help trainees to succeed in their professional journeys
and to design future generation telecommunication systems. Other institutions
that are willing to set up a similar training can profit from the
experiences shared in this article. Especially, the improvement of
trainees confirm the effectiveness and make the proposed training
a candidate for the flagship training of telecommunication education.

\section*{Acknowledgment}

The authors would like to thank former trainees: J. Olivo, A. Menon,
M. F. Kucuk, N. Soulandros, and G. Gillespie for their outstanding
projects that are presented in Fig. \ref{fig:Sample-Projects}. Also,
we would like to thank USF professors: R. D. Gitlin, N. Ghani, S.
Morgera, and Z. Lu for their courteously supplied academic success
statistics that are demonstrated in Table \ref{tab:Academic-Success-Comparison}.

\vspace{-0mm}

\bibliographystyle{IEEEtran}
\bibliography{IEEEabrv,WLAB_Refs}

\begin{thebibliography}{10}
\providecommand{\url}[1]{#1}
\csname url@samestyle\endcsname
\providecommand{\newblock}{\relax}
\providecommand{\bibinfo}[2]{#2}
\providecommand{\BIBentrySTDinterwordspacing}{\spaceskip=0pt\relax}
\providecommand{\BIBentryALTinterwordstretchfactor}{4}
\providecommand{\BIBentryALTinterwordspacing}{\spaceskip=\fontdimen2\font plus
\BIBentryALTinterwordstretchfactor\fontdimen3\font minus
  \fontdimen4\font\relax}
\providecommand{\BIBforeignlanguage}[2]{{%
\expandafter\ifx\csname l@#1\endcsname\relax
\typeout{** WARNING: IEEEtran.bst: No hyphenation pattern has been}%
\typeout{** loaded for the language `#1'. Using the pattern for}%
\typeout{** the default language instead.}%
\else
\language=\csname l@#1\endcsname
\fi
#2}}
\providecommand{\BIBdecl}{\relax}
\BIBdecl

\bibitem{parkvall2017}
S.~Parkvall, E.~Dahlman, A.~Furuskar, and M.~Frenne, ``{NR}: The new {5G} radio
  access technology,'' \emph{IEEE Communications Standards Magazine}, vol.~1,
  no.~4, pp. 24--30, Dec 2017.

\bibitem{consonni2010}
D.~Consonni and M.~T.~M. Silva, ``Signals in communication engineering
  history,'' \emph{{IEEE} Trans. Educ.}, vol.~53, no.~4, pp. 621--630, Nov
  2010.

\bibitem{OSI}
\emph{\BIBforeignlanguage{F}{{Information technology -- Open Systems
  Interconnection -- Network service definition}}}, International Organization
  for Standardization Standard ISO/IEC 8348:2002, Rev.~3, Nov. 2002.

\bibitem{padgett2006}
W.~T. Padgett, B.~A. Black, and B.~A. Ferguson, ``Low-frequency wireless
  communications system-infrared laboratory experiments,'' \emph{{IEEE} Trans.
  Educ.}, vol.~49, no.~1, pp. 49--57, Feb 2006.

\bibitem{cassara2006}
F.~A. Cassara, ``Wireless communications laboratory,'' \emph{{IEEE} Trans.
  Educ.}, vol.~49, no.~1, pp. 132--140, Feb 2006.

\bibitem{linn2012}
Y.~Linn, ``An ultra low cost wireless communications laboratory for education
  and research,'' \emph{{IEEE} Trans. Educ.}, vol.~55, no.~2, pp. 169--179, May
  2012.

\bibitem{aliakbarian2014}
H.~Aliakbarian, P.~J. Soh, S.~Farsi, H.~Xu, E.~H. E. M. J. C.~V. Lil, B.~K.
  J.~C. Nauwelaers, G.~A.~E. Vandenbosch, and D.~M. M.-P. Schreurs,
  ``Implementation of a project-based telecommunications engineering design
  course,'' \emph{{IEEE} Trans. Educ.}, vol.~57, no.~1, pp. 25--33, Feb 2014.

\bibitem{berber2018}
S.~M. Berber and K.~W. Sowerby, ``Visual presentation of abstract theoretical
  concepts using animations in communication systems courses,'' \emph{Computer
  Applications in Engineering Education}, vol.~26, no.~1, pp. 49--61, 2018.

\bibitem{guzelgoz-1-2010}
S.~Güzelgöz and H.~Arslan, ``A wireless communications systems laboratory
  course,'' \emph{{IEEE} Trans. Educ.}, vol.~53, no.~4, pp. 532--541, Nov 2010.

\bibitem{dale1969}
E.~Dale, \emph{Audio-visual methods in teaching}, 3rd~ed.\hskip 1em plus 0.5em
  minus 0.4em\relax New York: Holt, Rinehart \& Winston, 1969.

\bibitem{hoic-bozic2009}
N.~Hoic-Bozic, V.~Mornar, and I.~Boticki, ``A blended learning approach to
  course design and implementation,'' \emph{{IEEE} Trans. Educ.}, vol.~52,
  no.~1, pp. 19--30, Feb 2009.

\bibitem{guzelgoz-2-2010}
S.~Güzelgöz, S.~Yarkan, and H.~Arslan, ``Investigation of time selectivity of
  wireless channels through the use of {RVC},'' \emph{Measurement}, vol.~43,
  no.~10, pp. 1532--1541, 2010.

\bibitem{kihero2019}
A.~B. Kihero, M.~Karabacak, and H.~Arslan, ``Emulation techniques for small
  scale fading aspects by using reverberation chamber,'' \emph{{IEEE} Trans.
  Antennas Propag.}, vol.~67, no.~2, pp. 1246--1258, Feb 2019.

\bibitem{demir2017}
A.~F. Demir, Q.~H. Abbasi, Z.~E. Ankarali, A.~Alomainy, K.~Qaraqe, E.~Serpedin,
  and H.~Arslan, ``Anatomical region-specific in vivo wireless communication
  channel characterization,'' \emph{{IEEE} J. Biomed. Health Inform.}, vol.~21,
  no.~5, pp. 1254--1262, Sept 2017.

\bibitem{dogan2018}
S.~Dogan, M.~Karabacak, and H.~Arslan, ``Optimization of antenna beamwidth
  under blockage impact in millimeter-wave bands,'' in \emph{Proc. 2018 IEEE
  29th Annu. Int. Symp. Personal, Indoor and Mobile Radio Commun.}, Bologna,
  IT, Sep. 2018, pp. 1--5.

\end{thebibliography}

\vspace{-0mm}
\begin{IEEEbiography}[\includegraphics{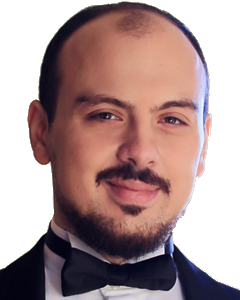}]{Ali Fatih Demir}
(S'08) received the B.S. degree in electrical engineering from Y\i ld\i z
Technical University, Istanbul, Turkey, in 2011 and the M.S. degrees
in electrical engineering and applied statistics from Syracuse University,
Syracuse, NY, USA in 2013. He is currently pursuing the Ph.D. degree
as a member of the Wireless Communication and Signal Processing (WCSP)
Group in the Department of Electrical Engineering, University of South
Florida, Tampa, FL, USA. His current research interests include PHY
and MAC aspects of wireless communication systems, \emph{in vivo}
wireless communication systems, and signal processing/machine learning
algorithms for brain-computer interfaces.
\end{IEEEbiography}

\vspace{-0mm}
\begin{IEEEbiography}[\includegraphics{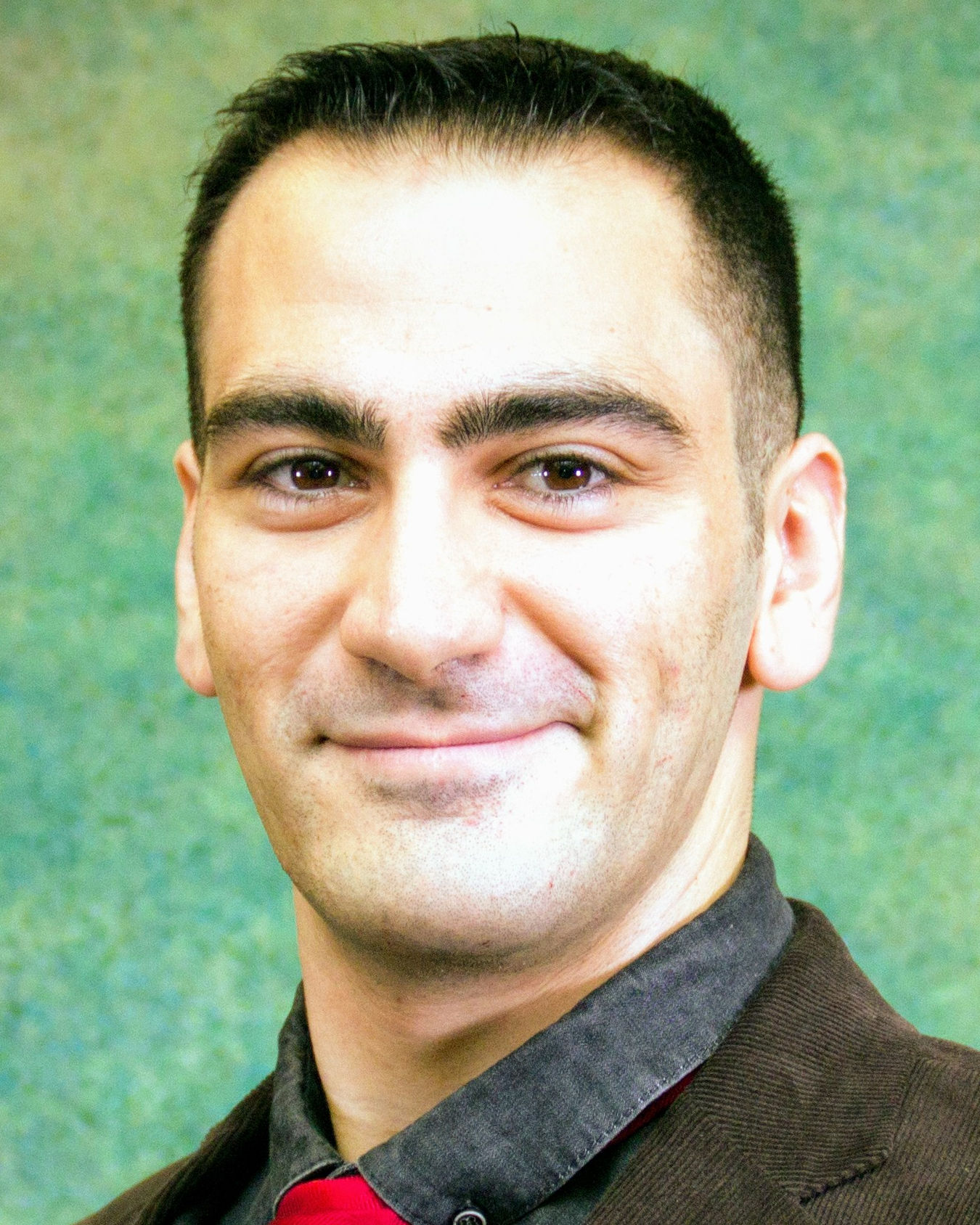}]{Berker Peköz}
 (GS\textquoteright 15) received the B.S. degree in electrical and
electronics engineering from Middle East Technical University, Ankara,
Turkey in 2015, and the M.S.E.E. from University of South Florida,
Tampa, FL, USA in 2017. He was a Co-op Intern at the Space Division
of Turkish Aerospace Industries, Inc., Ankara, Turkey in 2013, and
a Summer Intern at the Laboratory for High Performance DSP \& Network
Computing Research, New Jersey Institute of Technology, Newark, NJ,
USA in 2014. He is currently pursuing the Ph.D. at University of South
Florida, Tampa, FL, USA. His research is concerned with standard compliant
waveform design and optimization. Mr. Peköz is a member of Tau Beta
Pi.
\end{IEEEbiography}

\vspace{-0mm}
\begin{IEEEbiography}[\includegraphics{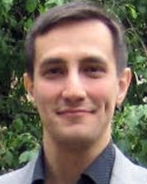}]{Selçuk Köse}
(S'10\textendash M\textquoteright 12) received the B.S. degree in
electrical and electronics engineering from Bilkent University, Ankara,
Turkey, in 2006, and the M.S. and Ph.D. degrees in electrical engineering
from the University of Rochester, Rochester, NY, USA, in 2008 and
2012, respectively. He was an Assistant Professor of Electrical Engineering
at the University of South Florida, Tampa, FL, USA. He is currently
an Associate Professor of Electrical and Computer Engineering at University
of Rochester, Rochester, NY, USA. His current research interests include
integrated voltage regulation, 3-D integration, hardware security,
and green computing. Dr. Köse was a recipient of the NSF CAREER Award,
the Cisco Research Award, the USF College of Engineering Outstanding
Junior Researcher Award, and the USF Outstanding Faculty Award. He
has served on the Technical Program and Organization Committees of
various conferences. He is currently an Associate Editor of the \emph{World
Scientific Journal of Circuits, Systems, and Computers} and the \emph{Elsevier
Microelectronics Journal}. 
\end{IEEEbiography}

\vspace{-0mm}
\begin{IEEEbiography}[\includegraphics{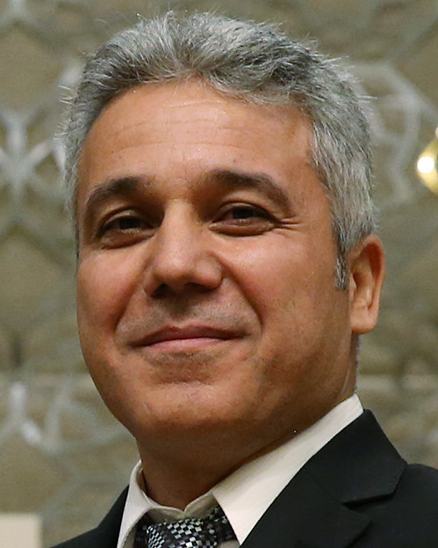}]{Hüseyin Arslan}
 (S\textquoteright 95\textendash M\textquoteright 98\textendash SM\textquoteright 04\textendash F\textquoteright 15)
received the B.S. degree in electrical and electronics engineering
from Middle East Technical University, Ankara, Turkey in 1992, and
the M.S. and Ph.D. degrees in electrical engineering from Southern
Methodist University, Dallas, TX, USA in 1994 and 1998, respectively.
From January 1998 to August 2002, he was with the research group of
Ericsson Inc., Charlotte, NC, USA, where he was involved with several
projects related to 2G and 3G wireless communication systems. He is
currently a Professor of Electrical Engineering at the University
of South Florida, Tampa, FL, USA, and the Dean of the College of Engineering
and Natural Sciences at the \.{I}stanbul Medipol University, \.{I}stanbul,
Turkey. His current research interests are on 5G and beyond, waveform
design, advanced multiple accessing techniques, physical layer security,
beamforming and massive MIMO, cognitive radio, dynamic spectrum access,
interference management (avoidance, awareness, and cancellation),
co-existence issues on heterogeneous networks, aeronautical (high
altitude platform) communications, millimeter-wave communications
and\emph{ in vivo} communications. He has served as technical program
committee chair, technical program committee member, session and symposium
organizer, and workshop chair in several IEEE conferences. He is currently
a member of the editorial board for the \emph{IEEE Communications
Surveys and Tutorials} and the \emph{Sensors Journal}.
\end{IEEEbiography}

\end{document}